\documentclass[11pt]{article}

\usepackage[final]{acl}

\usepackage{times}
\usepackage{latexsym}
\usepackage{amsfonts}
\usepackage{amsmath}
\usepackage[table]{xcolor}
\usepackage{dashrule}
\usepackage{enumitem}
\usepackage{listings}
\usepackage{booktabs}
\usepackage{multirow}

\usepackage[T1]{fontenc}

\usepackage[utf8]{inputenc}

\usepackage{microtype}

\usepackage{inconsolata}

\usepackage{graphicx}

\title{\textsc{TipCoder}: Reinforcement Learning Boosted \underline{T}est-time \underline{I}nstruction \underline{P}roposer for Code Generation}

\author{First Author \\
  Affiliation / Address line 1 \\
  Affiliation / Address line 2 \\
  Affiliation / Address line 3 \\
  \texttt{email@domain} \\\And
  Second Author \\
  Affiliation / Address line 1 \\
  Affiliation / Address line 2 \\
  Affiliation / Address line 3 \\
  \texttt{email@domain} \\}

\usepackage{tcolorbox}
\tcbuselibrary{skins, breakable}

\newtcolorbox{promptbox}[1][]{
    colback=gray!5!white,      
    colframe=gray!75!black,    
    title={\textbf{#1}},       
    fonttitle=\bfseries\small,
    fontupper=\small\ttfamily, 
    boxrule=0.8pt,
    arc=2pt,
    left=5pt, right=5pt, top=5pt, bottom=5pt,
    breakable,                 
}

\author{
\textbf{Minyu~Chen\textsuperscript{1}\thanks{~~Corresponding authors.},}
\textbf{Sihao~Wu\textsuperscript{2}},
\textbf{Ling-I~Wu\textsuperscript{3}},
\textbf{Song Qin\textsuperscript{1,4}},
\\
\textbf{Jingyang~Li\textsuperscript{3}},
\textbf{Lei~Ning\textsuperscript{1}},
\textbf{Jianxin~Xue\textsuperscript{4}},
\textbf{Guoqiang~Li\textsuperscript{3}\footnotemark[1]}
\\
\textsuperscript{1}Shenzhen Technology University, \\
\textsuperscript{2}Institute of AI for Industries, Chinese Academy of Sciences,\\
\textsuperscript{3}Shanghai Jiao Tong University,
\textsuperscript{4}Shanghai Polytechnic University\\
\texttt{chenminyu@sztu.edu.cn, li.g@sjtu.edu.cn}
}

\begin{document}
\maketitle

\begin{abstract}
Test-time scaling for code generation typically explores the solution space by sampling multiple programs from a fixed instruction. We study a complementary direction: instance-level instruction-space exploration. Our observation is that many coding failures stem from missing constraints, overlooked edge cases, or misleading reasoning paths induced by the original prompt. To address this, we propose \textsc{TipCoder}, a test-time instruction proposer that generates problem-specific auxiliary tips before code synthesis.
\textsc{TipCoder} distills multi-turn debugging trajectories into proactive guidance and further optimizes the Proposer with reinforcement learning using a marginal-utility reward. At inference time, it generates both a base solution and a tip-guided solution, and applies a Reward Model for post-hoc selection. This exploration-selection design allows tips to expose additional candidate potential while reducing regressions from unnecessary guidance. Across the evaluated code-generation benchmarks and target Code LLMs, \textsc{TipCoder} provides a consistent instruction-level test-time scaling strategy, comparing favorably with stochastic sampling and generic prompt optimization baselines under a shared reward-model-based selection protocol.
\begingroup
\renewcommand{\thefootnote}{\fnsymbol{footnote}}%
\footnote[2]{~~Our dataset and code are available at \url{https://github.com/Minkow/TipCoder}}
\endgroup
\end{abstract}

\section{Introduction}

Code Large Language Models~\cite{guo2024deepseek, hui2024qwen2} have become increasingly effective at translating natural-language specifications into executable programs. However, their outputs remain highly sensitive to the initial problem instruction, which may omit implicit constraints, under-specify boundary cases, or induce a spurious reasoning trajectory. Existing test-time scaling methods, such as Best-of-$N$ sampling~\cite{brown2024large}, improve reliability by allocating additional computation to explore the \textit{solution space}~\cite{snell2025scaling, muennighoff2025s1}. Since these methods typically sample from a fixed conditioning context, they may remain concentrated around the same flawed interpretation when the original instruction is misleading. This motivates a complementary scaling axis: adapting the instruction that guides solution search.

Prompt optimization demonstrates that instructions can be optimized to improve model behavior~\cite{zhou2022large, yang2023large, agrawalgepa, hu2025dipper}. For code generation, however, the limiting factor is often not the absence of a globally better prompt, but the absence of instance-specific guidance. A failure may hinge on a particular boundary case, index convention, library behavior, numerical constraint, or ambiguity in the specification. Such localized failure modes are poorly captured by fixed dataset-level prompts or generic reminders such as ``consider edge cases''. We therefore study \textit{instance-level instruction proposing}: generating a problem-specific auxiliary instruction before code synthesis.

We introduce \textsc{TipCoder}, a test-time instruction proposer for code generation. Given a programming problem, a \texttt{Proposer} generates an auxiliary tip that highlights relevant constraints, likely pitfalls, or implementation considerations. The target Code LLM then produces two candidates: a base solution conditioned on the original instruction and a tip-guided solution conditioned on the augmented instruction. The tip is not intended to specify the solution; instead, it modifies the conditioning context so that the generator explores an alternative region of the solution space.

To learn such tips, \textsc{TipCoder} distills multi-turn debugging trajectories into proactive guidance. In these trajectories, an initial solution fails, feedback exposes a missing requirement or incorrect assumption, and a subsequent solution repairs the error. This process reveals information that would have been useful before the initial generation. We first initialize the Proposer with Supervised Fine-Tuning on distilled repair guidance, and then optimize it with RL in a frozen Code LLM environment. The reward measures the marginal utility of the generated tip relative to base generation, encouraging corrective instructions while penalizing harmful or redundant perturbations.

At inference time, the utility of a tip is instance-dependent. A tip-guided candidate may recover a solution that is unlikely under the original prompt, but can also perturb an already correct base trajectory. Thus, the base and tip-guided candidates define an oracle-selection upper bound for instruction-guided scaling, while practical deployment requires instance-wise selection. \textsc{TipCoder} uses a Reward Model (RM) for this post-hoc selection. The RM does not create instruction-level exploration, but approximates the selector needed to realize the candidate potential exposed by the Proposer.

\textsc{TipCoder} operates as an external prompt-level module and requires no access to the target Code LLM's parameters, gradients, or execution feedback during inference. Across multiple code-generation benchmarks and backbones, \textsc{TipCoder} consistently improves pass rates over base generation and stochastic sampling. Under the same RM reranking protocol, it also outperforms self-hinting and generic context-optimization baselines, indicating that the gains arise from more effective instance-level instruction proposals rather than from the mere availability of an additional candidate selector.

Our contributions are summarized as follows:
\begin{itemize}
    \item \textbf{Instance-level instruction-space exploration.}
    We formulate test-time instruction proposing as a complementary scaling axis for code generation, where problem-specific auxiliary instructions guide solution search beyond fixed-prompt stochastic sampling and dataset-level prompt optimization.

    \item \textbf{Debugging-distilled and utility-optimized Proposer.}
    We distill multi-turn debugging trajectories into proactive tips that capture latent constraints, edge cases, and failure modes. We then optimize the Proposer with RL using a marginal-utility reward.

    \item \textbf{Exploration-selection decomposition.}
    We use RM-based reranking to approximate instance-wise selection between base and tip-guided candidates, and evaluate \textsc{TipCoder} against stochastic sampling, self-hinting, and context-optimization baselines under the same reranking protocol.
\end{itemize}

\section{Methodology}

\begin{figure*}[t]
    \centering
    \includegraphics[width=0.98\textwidth]{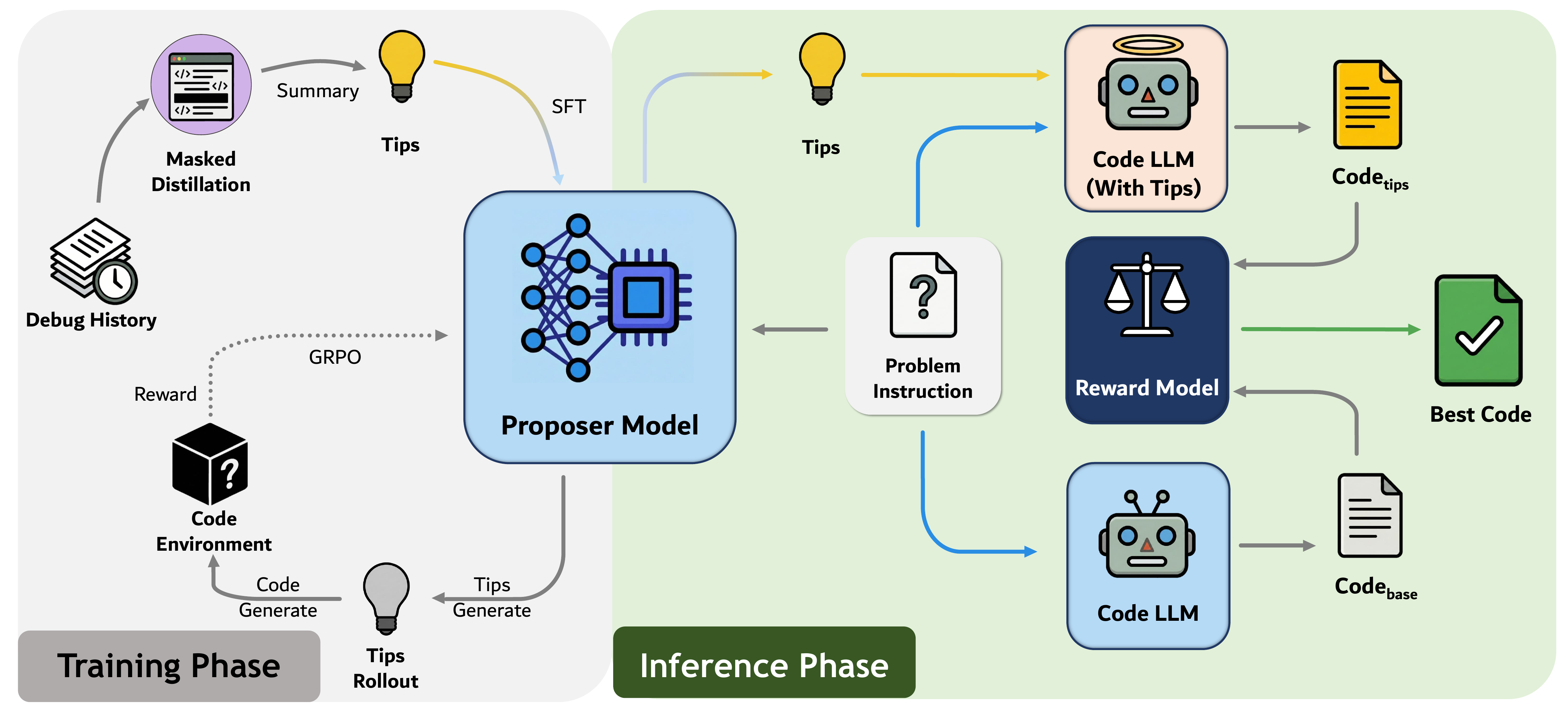}
    \caption{Overview of \textsc{TipCoder}. The training phase (left) consists of two stages: initializing the Proposer Agent via SFT, followed by Reinforcement Learning using pass rates from the code environment. The inference phase (right) employs the Proposer Model to generate guiding tips, and a Reward Model is used to select the best solution between the base and guided codes.}
    \label{fig:framework}
\end{figure*}

\subsection{Problem Formulation}

Let $\mathcal{D} = \{(x, T)\}$ denote a dataset where $x$ is the problem description and $T$ is the unit test set. A Code LLM $\mathcal{M}$ generates a solution $y$ according to $P(y|x) = \mathcal{M}(y|x)$.
While standard one-off generation often fails due to ambiguous constraints, empirical evidence suggests that a correct solution $y^*$ remains reachable \cite{austin2021program, larbi2024ambiguity}. This can be achieved through iterative debugging with execution feedback, where error messages are utilized to refine the code over multiple turns \cite{chen2023self, shinn2023reflexion}.

Our goal is to retain useful signals from this multi-turn process without requiring a runtime debugging loop. We posit that the sparse reasoning signals within the interaction history can be compressed into a concise latent instruction $c$ (a ``tip''). Consequently, we formulate the problem as learning a proposer policy $\pi_\phi(c|x)$ that maximizes the expected pass rate of the downstream generation:
\begin{equation}
\begin{split}
    \max_{\phi} \mathcal{J}(\phi) = \mathbb{E}_{c \sim \pi_\phi(\cdot|x)} \Big[ & \mathbb{E}_{y \sim \mathcal{M}(\cdot|x, c)} \\
    & [\mathbb{1}(\text{Pass}(y, T))] \Big]. \nonumber
\end{split}
\end{equation}
Intuitively, this objective drives the proposer to discover optimal shortcuts in the instruction space that functionally replicate the utility of the expensive debugging history.

\subsection{Tip Supervision from Debugging Trajectories}

Training the Proposer requires high-quality pairs $(x,c)$ where the auxiliary instruction $c$ captures information that is missing from the original problem instruction but useful for code generation. We construct such supervision by distilling successful multi-turn debugging trajectories into single-turn proactive tips. The construction pipeline consists of three stages: mining repair trajectories, extracting debugging observations through masked distillation, and verifying the causal utility of the tips.

\paragraph{Mining Repair Trajectories.}
We use AceCoder-87K~\cite{acecoder}, a large-scale code instruction dataset constructed through evolutionary instruction generation from open-source repositories, as the seed corpus. To obtain dense reasoning signals, we focus on challenging samples where the student model exhibits low one-pass performance, filtering for problems with Pass@1 $<50\%$. For each retained problem, we employ Gemini-3-Pro as the teacher model to perform iterative debugging with execution feedback.

For a problem $x$ with unit tests $T$, we denote a debugging trajectory as
\begin{equation}
    \tau = \{(y^0, e^0), (y^1, e^1), \ldots, (y^K, e^K)\},
\end{equation}
where $y^t$ is the solution generated at turn $t$ and $e^t$ is the corresponding execution feedback. We retain trajectories where the initial solution fails but is successfully repaired within three turns:
\begin{equation}
    \mathrm{Pass}(y^0,T)=0,\qquad \mathrm{Pass}(y^K,T)=1.
\end{equation}
Such trajectories capture the transition from an incorrect initial interpretation to a correct solution, and therefore expose the information that the original instruction failed to elicit.

\paragraph{Abductive Reasoning via Masked Distillation.}
A successful repair trajectory contains a \emph{debugging observation}: the hidden constraint, edge case, or invalid assumption that explains why the initial solution failed and how the later solution repaired it. The goal of distillation is to convert this retrospective observation into a proactive instruction that can be provided before code generation. In this sense, the tip serves as a compact substitute for the foresight normally obtained only after costly iterative debugging.

To prevent solution leakage, we apply masked distillation. Implementation-level code blocks and solution-specific details are masked before the teacher extracts the observation. Formally, the teacher first infers the debugging observation from the problem and the masked trajectory:
\begin{equation}
    o_\tau = \mathrm{Obs}_{\mathrm{teacher}}(x,\mathrm{Mask}(\tau)).
\end{equation}
The observation is then verbalized as an auxiliary tip:
\begin{equation}
    c \sim q_{\mathrm{teacher}}(\cdot \mid x,o_\tau).
\end{equation}
The generated tip is required to describe high-level constraints, edge cases, or likely failure modes rather than prescribe the final implementation. This process distills the retrospective knowledge revealed by multi-turn repair into a single-turn guidance signal available at the initial generation stage.

\paragraph{Iterative Feedback Verification.}
The distilled tip is retained only if it is causally useful for the student model. We append the generated tip $c$ to the original instruction $x$ and run the student model again. If the generated code still fails, the execution feedback is returned to the teacher model to refine the tip, with at most two refinement rounds. A pair $(x,c)$ is added to the final training set if and only if the verified tip guides the student model to pass all unit tests:
\begin{equation}
    \mathrm{Pass}(y_{\mathrm{tip}},T)=1.
\end{equation}
This filtering process yields 2,126 high-quality problem-tip pairs for supervised initialization.

\subsection{Proposer Optimization}

We optimize the Proposer in two stages. Supervised Fine-Tuning initializes the policy with verified debugging-derived tips, and Reinforcement Learning further aligns the policy with downstream code correctness.

\paragraph{Supervised Fine-Tuning (SFT).}
We initialize the Proposer policy $\pi_\phi$ using the distilled dataset $\mathcal{D}_{train}$. This stage teaches the Proposer the form and patterns of useful tips. However, it does not directly optimize downstream correctness, motivating the subsequent RL stage.

\paragraph{Latent-Variable Interpretation.}
The tip $c$ mediates the relationship between the problem $x$ and a correct solution $y^*$. SFT supplies a reference distribution over natural-language tips, and RL searches for tips with higher downstream utility.

Taking the frozen SFT model $\pi_{\text{ref}}(c|x)$ as a prior gives $P(y^*|x)=\sum_c P(y^*|x,c)\pi_{\text{ref}}(c|x)$. Using the Proposer policy $\pi_\phi(c|x)$ as a variational distribution yields the following Evidence Lower Bound:

\begin{equation}
\resizebox{\linewidth}{!}{$
\begin{aligned}
\log P(y^*|x) & = \log \sum_{c} \pi_\phi(c|x) \frac{P(y^*|x, c) \pi_{\text{ref}}(c|x)}{\pi_\phi(c|x)} \\
& \ge \sum_{c} \pi_\phi(c|x) \log \frac{P(y^*|x, c) \pi_{\text{ref}}(c|x)}{\pi_\phi(c|x)} \\
& = \underbrace{\mathbb{E}_{c \sim \pi_\phi} [\log P(y^*|x, c)]}_{\text{Reconstruction Objective}} - \underbrace{D_{KL}(\pi_\phi(\cdot|x) || \pi_{\text{ref}}(\cdot|x))}_{\text{Regularization Constraint}}.\nonumber
\end{aligned}$
}
\end{equation}

Here $P(y^*|x,c)$ denotes the likelihood of the fixed Code LLM generating a correct solution given the tip. The implemented objective uses execution pass rates, the piecewise marginal-utility reward below, and a length penalty as operational approximations to the reconstruction term.

\paragraph{Reinforcement Learning Objective.}
Since $P(y^*|x,c)$ is not directly accessible in code generation, we replace it with an execution-based reward. We define the instruction proposer $\pi_\phi(c|x)$ as the \textbf{policy} and the combination of the frozen Code LLM and the deterministic execution sandbox as the \textbf{environment}. The environment receives a tip $c$, generates code, and returns a non-differentiable reward based on execution results.

The optimization objective $\mathcal{J}(\phi)$ maximizes the expected execution-based reward with KL regularization:
\begin{equation}
\begin{aligned}
\mathcal{J}(\phi) = & \mathbb{E}_{c \sim \pi_\phi(\cdot|x)} [R(c)] \\
& - \beta D_{KL}(\pi_\phi(\cdot|x) || \pi_{\text{ref}}(\cdot|x)). \nonumber
\end{aligned}
\end{equation}
We optimize this objective with Group Relative Policy Optimization (GRPO)~\cite{shao2024deepseekmath}. For each input $x$, we sample a group of outputs $\{c_1, \dots, c_G\}$ and estimate the baseline using the group average.

\paragraph{Marginal-Utility Reward.}
A naive reward based only on the pass rate of the tip-guided solution is noisy: on simple tasks where the base model already succeeds, the policy can receive high reward even for redundant or empty tips. We therefore define the reward by the marginal utility of the tip relative to base generation.

Let $P_{\text{base}}$ denote the pass rate of the code generated with the original problem description and $P_{\text{curr}}$ denote the pass rate with the generated tip $c$. We define the Utility Gain $\Delta$ to explicitly measure the improvement brought by the tip:
\begin{equation}
    \Delta =
    \begin{cases}
    1.0 & \text{if } P_{\text{base}} < 1 \text{ and } P_{\text{curr}} = 1 \\
    P_{\text{curr}} - P_{\text{base}} & \text{otherwise}
    \end{cases}\nonumber
\end{equation}

This reward approximates the marginal contribution of the proposed tip before RM-based selection. A positive value indicates that the tip improves the tip-guided branch relative to the base branch, while a negative value indicates regression. Subtracting the base performance filters out redundant tips on already solved tasks and makes the reward depend on the relative contribution of the instruction rather than absolute task difficulty. This centering also reduces reward variance, encouraging policy updates to focus on whether the tip changes the outcome.

Finally, to further encourage conciseness, we incorporate a length penalty using a ReLU style threshold. The total reward $R(c)$ is defined as
\begin{equation}
    R(c) = \Delta - \lambda \cdot \max(0, \mathcal{L}(c) - \mathcal{L}_{\text{th}}), \nonumber
\end{equation}
where $\mathcal{L}(c)$ is the token length of the tip and $\mathcal{L}_{\text{th}}=2000$ is the soft threshold.

\subsection{Reward-Guided Reranking Inference}
\label{sec:rerank_inference}

At inference time, \textsc{TipCoder} follows a generate-and-rerank protocol. The Proposer generates a problem-specific tip $c$ for every input problem $x$. We do not require the Proposer to decide whether the base model would fail on the original prompt, since problem difficulty and failure modes are model-dependent. Instead, \textsc{TipCoder} constructs two branches: the base branch from the original instruction and the tip-guided branch from the augmented instruction.

\begin{itemize}
    \item \textbf{Base branch:} the target Code LLM generates $y_{\mathrm{base}}$ from the original instruction $x$.
    \item \textbf{Tip-guided branch:} the same target Code LLM generates $y_{\mathrm{tip}}$ from the augmented instruction $x \oplus c$.
\end{itemize}

\textsc{TipCoder} then applies a Reward Model $R$ as a post-hoc selector:
\begin{equation}
    \hat{y} =
    \begin{cases}
    y_{\mathrm{tip}}, & R(x,y_{\mathrm{tip}}) > R(x,y_{\mathrm{base}}), \\
    y_{\mathrm{base}}, & \mathrm{otherwise}.
    \end{cases}
\end{equation}
The deployed objective is therefore to improve the pass rate of the selected output,
\begin{equation}
    \max_\phi\;
    \mathbb{E}_{c \sim \pi_\phi(\cdot \mid x)}
    \mathbb{E}_{y_{\mathrm{base}}, y_{\mathrm{tip}}}
    \left[
    \mathbb{1}\big(\mathrm{Pass}(\hat{y},T)\big)
    \right].
\end{equation}


This protocol completes the exploration-selection decomposition of \textsc{TipCoder}: the Proposer exposes an alternative instruction-conditioned candidate, and the Reward Model selects between the base and tip-guided solutions for each instance. This post-hoc selection reduces regressions from unnecessary or noisy tips without requiring the Proposer to predict whether the base model will fail.

\textsc{TipCoder} is black-box with respect to the target Code LLM. It operates only through prompt-level augmentation and candidate scoring, requiring no access to model parameters, gradients, or execution feedback during inference. Because it explores the instruction dimension rather than replacing solution-level sampling, it remains compatible with test-time scaling methods such as Best-of-$N$.

\section{Experiments}

\subsection{Setup}

\noindent \textbf{Datasets and Metrics.}
We evaluate on three code-generation benchmarks: HumanEval+~\cite{chen2021evaluating}, MBPP+~\cite{austin2021program}, and BigCodeBench-Instruct~\cite{zhuobigcodebench}. For HumanEval+ and MBPP+, we use the EvalPlus implementation~\cite{liu2023your}, which provides more rigorous test suites than the original benchmarks. HumanEval and MBPP primarily cover algorithmic and entry-level Python programming tasks, while BigCodeBench-Instruct evaluates more realistic programming scenarios involving library usage and complex function calls. Since \textsc{TipCoder} targets instruction refinement, we evaluate on the instruction-following split of BigCodeBench rather than the completion split. We use Pass@1 as the primary metric. Evaluation tests are used only for reporting final correctness and are not available to \textsc{TipCoder} during inference.

\noindent \textbf{Proposer Training.}
The Proposer is initialized from \texttt{Qwen3-4B-Instruct-2507}. We use a general instruction model rather than a code-specialized model because tip generation requires identifying constraints, edge cases, and failure modes in natural language. During RL, the Proposer interacts with a frozen code-generation environment instantiated by \texttt{Qwen2.5-7B-Coder-Instruct}. This environment is used only for training-time rollouts and reward computation; it is not the sole target model used in evaluation.

For RL training, we use the hard subset of AceCoder-87K and exclude all samples used in the SFT stage. We further downsample extreme cases where the base pass rate is either 0 or 1, so that the policy receives informative marginal-utility signals rather than learning from tasks that are already solved or consistently unsolved by the rollout environment. This yields 13,376 RL training samples. The evaluation benchmarks are held out from Proposer training.

We implement training with OpenRLHF and vLLM. The rollout group size is set to 8. The environment Code LLM uses temperature 0 to provide deterministic execution feedback during RL, while the Proposer samples tips with temperature 1.0 during training rollouts and uses greedy decoding during evaluation. Full hyperparameters are provided in Appendix~\ref{app:hyperparams}.

\noindent \textbf{Target Code LLMs.}
We evaluate \textsc{TipCoder} on four open-source Code LLM backbones: \texttt{Qwen2.5-Coder-7B-Instruct}, \texttt{DeepSeek-Coder-7B-Instruct-v1.5}, \texttt{Qwen3-Coder-30B-A3B-Instruct}, and \texttt{DeepSeek-Coder-V2-Lite-Instruct}. Unless otherwise specified, the same trained Proposer is applied to all target backbones without target-specific fine-tuning. This setting tests whether the learned instruction proposer transfers across different Code LLMs.

\definecolor{sftgray}{RGB}{245, 245, 245}
\definecolor{rlblue}{RGB}{225, 245, 254}

\newcommand{\ms}[2]{#1{\scriptsize$\pm$#2}}
\newcommand{\bms}[2]{\textbf{#1}{\scriptsize$\pm$#2}}
\newcommand{\sftc}[1]{\cellcolor{sftgray}#1}
\newcommand{\rlc}[1]{\cellcolor{rlblue}#1}

\setlength{\aboverulesep}{0pt}
\setlength{\belowrulesep}{0pt}
\renewcommand{\arraystretch}{1.12}

\begin{table*}[!ht]
    \centering
\caption{
Main results on code-generation benchmarks across diverse backbones.
Results are reported as mean{\scriptsize$\pm$}std over three independent runs.
All methods that require candidate selection use the same \texttt{AceCodeRM-32B} Reward Model.
Avg. denotes the unweighted average of the three benchmark means.
}
    \label{tab:main_results}
    \small
    \setlength{\tabcolsep}{3.6mm}

    \begin{tabular}{llcccc}
    \toprule
    \textbf{Model} & \textbf{Method} & \textbf{HumanEval+} & \textbf{MBPP+} & \textbf{BigCodeBench} & \textbf{Avg.} \\
    \midrule

    \multirow{8}{*}{\textbf{DeepSeek-Coder-7B}}
        & Base            & \ms{62.40}{1.86} & \ms{61.55}{2.12} & \ms{37.43}{0.05} & 53.79 \\
        & Best-of-2       & \ms{69.92}{1.86} & \ms{66.93}{2.38} & \ms{32.54}{0.00} & 56.46 \\
        & OPRO            & \ms{65.85}{0.00} & \ms{68.25}{0.00} & \ms{39.80}{0.05} & 57.97 \\
        & GEPA            & \ms{65.85}{0.00} & \ms{68.25}{0.00} & \ms{39.94}{0.05} & 58.01 \\
        & CTRL            & \ms{67.89}{0.93} & \ms{68.96}{0.15} & \ms{38.39}{0.48} & 58.41 \\
        & Self-Hint     & \ms{69.92}{1.86} & \ms{69.40}{0.31} & \ms{38.77}{0.61} & 59.36 \\
        & \sftc{\textsc{TipCoder}-SFT}  & \sftc{\ms{71.34}{0.00}} & \sftc{\ms{71.08}{0.81}} & \sftc{\ms{39.44}{0.05}} & \sftc{60.62} \\
        & \rlc{\textsc{TipCoder}-RL} & \rlc{\bms{72.36}{1.76}} & \rlc{\bms{71.16}{1.15}} & \rlc{\bms{40.00}{0.05}} & \rlc{\textbf{61.17}} \\
    \addlinespace[0.55em]

    \multirow{8}{*}{\textbf{DeepSeek-Coder-V2-Lite}}
        & Base            & \ms{74.39}{1.61} & \ms{68.78}{1.40} & \ms{38.19}{0.33} & 60.45 \\
        & Best-of-2       & \ms{78.25}{0.93} & \ms{71.34}{1.00} & \ms{34.21}{0.00} & 61.27 \\
        & OPRO            & \ms{76.22}{1.61} & \ms{72.13}{0.61} & \ms{40.88}{0.09} & 63.08 \\
        & GEPA            & \ms{77.03}{0.35} & \ms{72.49}{0.00} & \ms{40.61}{0.69} & 63.38 \\
        & CTRL            & \ms{78.66}{0.00} & \ms{72.93}{0.40} & \ms{39.04}{0.55} & 63.54 \\
        & Self-Hint     & \ms{80.49}{1.61} & \ms{71.60}{0.15} & \ms{40.96}{1.14} & 64.35 \\
        & \sftc{\textsc{TipCoder}-SFT} & \sftc{\ms{78.66}{0.61}} & \sftc{\ms{72.13}{1.10}} & \sftc{\ms{39.82}{0.23}} & \sftc{63.54} \\
        & \rlc{\textsc{TipCoder}-RL}  & \rlc{\bms{80.89}{1.04}} & \rlc{\bms{73.46}{0.31}} & \rlc{\bms{41.46}{0.43}} & \rlc{\textbf{65.27}} \\
    \addlinespace[0.55em]

    \multirow{8}{*}{\textbf{Qwen2.5-Coder-7B}}
        & Base            & \ms{78.86}{1.27} & \ms{64.46}{1.00} & \ms{41.64}{0.05} & 61.65 \\
        & Best-of-2       & \ms{85.77}{1.27} & \ms{70.37}{0.92} & \ms{36.32}{0.00} & 64.15 \\
        & OPRO            & \ms{85.57}{0.35} & \bms{73.81}{0.00} & \ms{43.57}{0.05} & 67.65 \\
        & GEPA            & \ms{85.57}{0.35} & \bms{73.81}{0.00} & \ms{42.87}{0.05} & 67.42 \\
        & CTRL            & \ms{85.57}{0.35} & \ms{73.54}{0.46} & \ms{42.08}{0.27} & 67.06 \\
        & Self-Hint     & \ms{86.38}{0.35} & \ms{73.54}{0.46} & \ms{41.49}{0.00} & 67.14 \\
        & \sftc{\textsc{TipCoder}-SFT} & \sftc{\ms{84.96}{0.35}} & \sftc{\ms{73.19}{1.07}} & \sftc{\ms{42.34}{0.05}} & \sftc{66.83} \\
        & \rlc{\textsc{TipCoder}-RL}  & \rlc{\bms{88.21}{0.35}} & \rlc{\bms{73.81}{0.92}} & \rlc{\bms{43.89}{0.56}} & \rlc{\textbf{68.64}} \\
    \addlinespace[0.55em]

    \multirow{8}{*}{\textbf{Qwen3-Coder-30B-A3B}}
        & Base            & \ms{86.99}{0.93} & \ms{73.72}{0.40} & \ms{48.39}{0.05} & 69.70 \\
        & Best-of-2       & \ms{87.60}{0.70} & \ms{74.60}{1.15} & \ms{43.16}{0.00} & 68.45 \\
        & OPRO            & \ms{89.02}{0.35} & \ms{74.07}{0.46} & \ms{48.71}{0.20} & 70.60 \\
        & GEPA            & \ms{89.02}{0.35} & \ms{74.34}{0.00} & \ms{48.54}{0.18} & 70.63 \\
        & CTRL            & \ms{87.80}{0.00} & \ms{73.46}{0.31} & \ms{48.22}{0.28} & 69.83 \\
        & Self-Hint     & \ms{88.62}{1.27} & \ms{73.90}{0.40} & \ms{46.29}{0.37} & 69.60 \\
        & \sftc{\textsc{TipCoder}-SFT} & \sftc{\ms{89.02}{0.70}} & \sftc{\ms{74.96}{0.15}} & \sftc{\ms{47.98}{0.30}} & \sftc{70.65} \\
        & \rlc{\textsc{TipCoder}-RL}  & \rlc{\bms{89.23}{0.93}} & \rlc{\bms{75.31}{0.33}} & \rlc{\bms{48.77}{0.09}} & \rlc{\textbf{71.10}} \\
    \bottomrule
    \end{tabular}
\end{table*}

\noindent \textbf{Baselines.}
We compare against the following baselines.
\textbf{Base} directly generates code from the original instruction.
\textbf{Best-of-$N$} samples multiple candidate solutions from the original instruction. In practice, we set $N=2$.
\textbf{Self-Hint} is a zero-shot instruction-proposing baseline. It uses the same tip-generation prompt as \textsc{TipCoder}, but replaces the trained Proposer with an untrained instruction model to generate the auxiliary tip; the target Code LLM then generates a tip-guided candidate from the resulting augmented instruction.
To compare with generic prompt or context optimization, we include
\textbf{OPRO}~\cite{yang2023large}, which treats prompt design as an optimization problem and iteratively proposes improved prompts;
\textbf{GEPA}~\cite{agrawalgepa}, which evolves prompts through reflective feedback; and
\textbf{CTRL-32B}~\cite{xie2025teaching}, a critique-based model that produces contextual guidance for code revision.

\noindent \textbf{Reward Model.}
For \textbf{all baselines} and \textsc{TipCoder}, the candidate set consists of the base solution and the corresponding augmented solution. We use the same \texttt{AceCodeRM-32B}~\cite{acecoder} to select the best candidate, without access to execution feedback or hidden tests at inference time. This keeps the selector fixed across methods, so performance differences reflect the generated candidate set rather than the reward model.

\begin{figure*}[t]
    \centering
    \includegraphics[width=\textwidth]{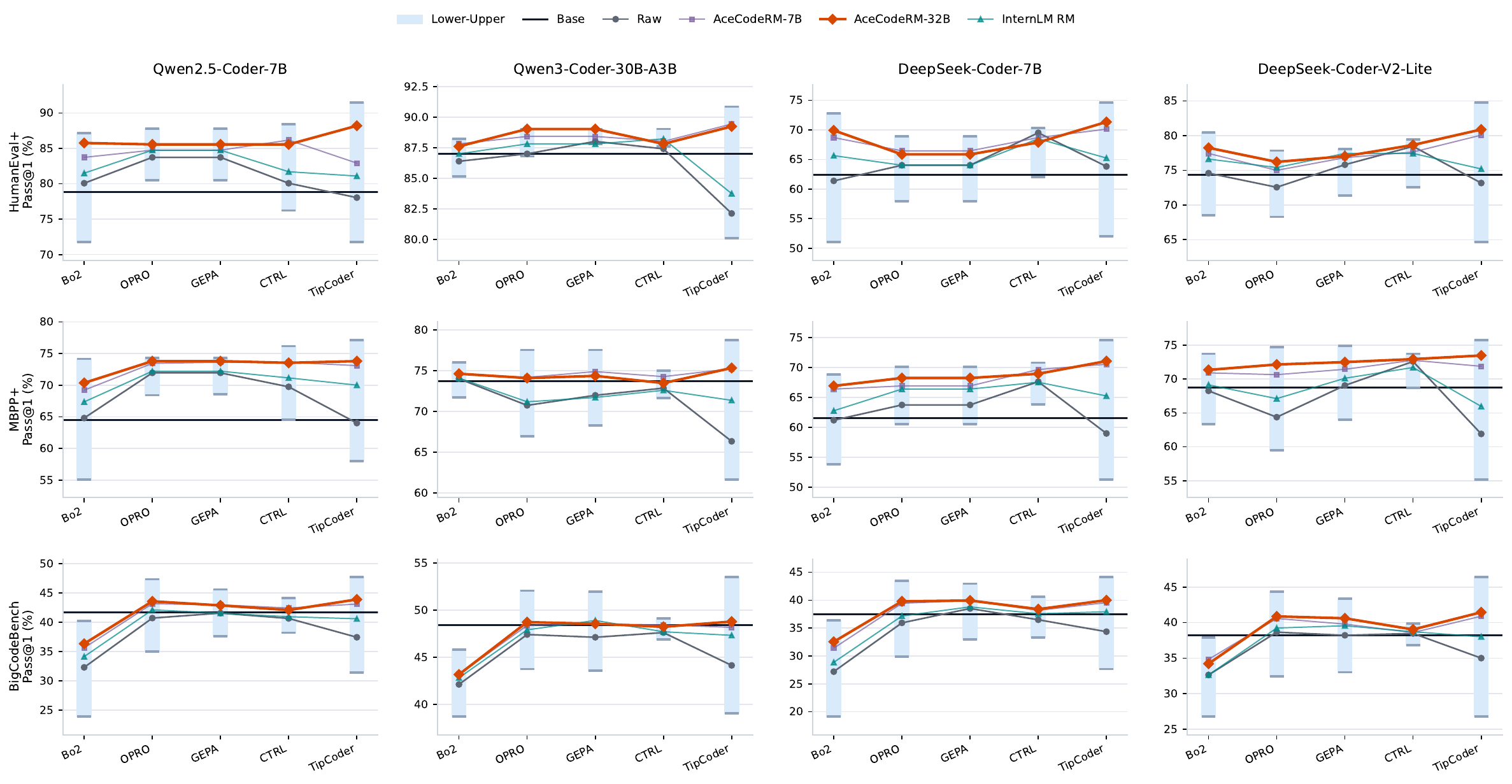}
    \caption{
    Disentangling candidate exploration from post-hoc selection.
    For each target model and benchmark, we fix the candidate set produced by each method and vary only the selection rule.
    The figure explicitly reports base performance, raw adoption of the additional branch, reward-model-selected performance under different Reward Models, and oracle upper/lower bounds over the candidate set.
    The shaded region denotes the oracle lower-to-upper range, where the upper bound measures the best achievable performance with an ideal selector and the lower bound measures the risk of selecting the worse candidate.
    A higher oracle upper bound indicates that a method exposes additional correct candidates, while the gap between RM-selected performance and the oracle upper bound reflects the remaining selection bottleneck.
    }
    \label{fig:bounds}
\end{figure*}

\subsection{Main Results}

Table~\ref{tab:main_results} reports the main results across four target Code LLMs. All methods requiring candidate selection use the same \texttt{AceCodeRM-32B} Reward Model. Prompt- and context-level optimization generally improves over base generation, but the gains vary across backbones and benchmarks. OPRO and GEPA are competitive on \texttt{Qwen2.5-Coder-7B} and \texttt{Qwen3-Coder-30B-A3B}, but are less stable on \texttt{DeepSeek-Coder-7B}. Best-of-2 improves HumanEval+ and MBPP+ in many cases, yet consistently degrades BigCodeBench, suggesting that merely sampling from the original instruction is insufficient for more realistic programming tasks. Table~\ref{tab:cost_normalized_bon} compares a range of Best-of-$N$ budgets, and Appendix~\ref{app:best_of_n_mbpp} extends the MBPP+ scaling curves to all four target backbones.

\textsc{TipCoder}-RL achieves the best average performance on all four backbones, reaching 61.17 on \texttt{DeepSeek-Coder-7B}, 65.27 on \texttt{DeepSeek-Coder-V2-Lite}, 68.64 on \texttt{Qwen2.5-Coder-7B}, and 71.10 on \texttt{Qwen3-Coder-30B-A3B}. Compared with the strongest non-\textsc{TipCoder} baseline on each backbone, this yields average improvements of 1.81, 0.92, 0.99, and 0.47 points, respectively. These results indicate that instance-level instruction proposing provides a more stable scaling direction than generic prompt optimization.

Self-Hint uses the same tip-generation prompt without training the Proposer, serving as a zero-shot instruction-refinement baseline. The trained \textsc{TipCoder} variants consistently improve over Self-Hint in average performance, showing that the gains come not merely from adding auxiliary instructions, but from learning problem-specific tips aligned with downstream code correctness.

\subsection{Disentangling Exploration from Selection}
\label{sec:exploration_selection}

Figure~\ref{fig:bounds} decomposes test-time performance into two factors: candidate exploration and post-hoc selection.
For each method, the figure explicitly reports the base performance, raw adoption of the additional branch, reward-model-selected performance under different Reward Models, and the oracle upper/lower bounds over the candidate set.
The oracle upper bound measures the best achievable performance if an ideal selector could choose between the base and additional candidates, while the lower bound measures the worst possible selection.
Thus, the gap between base performance and the oracle upper bound reflects the exploration potential introduced by the method, whereas the gap between reward-model selection and the oracle upper bound reflects how much of this potential remains unrealized by the selector.

The results show that instruction-guided generation can expose useful candidates beyond the original prompt.
In many settings, \textsc{TipCoder}-RL raises the oracle upper bound, indicating that learned tips create complementary solution paths rather than merely perturbing the prompt.
However, this potential is not automatically realized: raw adoption can select harmful candidates, and weaker selectors may fail to recover the improved branch.
Stronger code-oriented Reward Models, such as \texttt{AceCodeRM-32B}, generally convert more of the oracle potential into final pass-rate gains than weaker or less aligned selectors.
Therefore, the Reward Model is not the source of the exploration benefit; it determines how much of the candidate-space improvement opened by the Proposer can be realized at inference time.

\subsection{Cost-Normalized Best-of-$N$ Scaling}
\label{sec:cost_normalized_bon}

We compare Best-of-$N$ under a controlled test-time budget, holding the target Code LLM, original prompt, and \texttt{AceCodeRM-32B} selector fixed while varying $N\in\{1,2,4,8,16\}$. Table~\ref{tab:cost_normalized_bon} reports Pass@1, token counts, and wall-clock latency for Qwen2.5-Coder-7B on all three benchmarks.

\begin{table*}[t]
    \centering
    \small
    \caption{Cost-normalized comparison on \texttt{Qwen2.5-Coder-7B}. For Best-of-$N$, tokens include one shared coder prefill, $N$ code completions, and $N$ RM inputs (no RM for $N=1$); \textsc{TipCoder} includes one Proposer call, two code branches, and two RM inputs. Wall-clock latency is measured on a fixed 100-problem subset of each benchmark using one NVIDIA A800, excluding model loading. For $N>1$, latency is code generation plus RM scoring; \textsc{TipCoder} latency is $\max(\text{base code},\text{proposer}+\text{guided code})+\text{RM}$.}
    \label{tab:cost_normalized_bon}
    \resizebox{\textwidth}{!}{
    \begin{tabular}{lrrrrrrrrr}
        \toprule
        & \multicolumn{3}{c}{\textbf{HumanEval+}} & \multicolumn{3}{c}{\textbf{MBPP+}} & \multicolumn{3}{c}{\textbf{BigCodeBench-full}} \\
        \cmidrule(lr){2-4}\cmidrule(lr){5-7}\cmidrule(lr){8-10}
        \textbf{Method} & \textbf{Pass@1} & \textbf{Tokens} & \textbf{Wall (s)} & \textbf{Pass@1} & \textbf{Tokens} & \textbf{Wall (s)} & \textbf{Pass@1} & \textbf{Tokens} & \textbf{Wall (s)} \\
        \midrule
        Best-of-1 & 78.86 & 482.65 & 0.36 & 64.46 & 250.47 & 0.22 & 41.64 & 511.76 & 0.49 \\
        Best-of-2+RM & 85.77 & 1,322.29 & 0.77 & 70.37 & 712.30 & 0.47 & 36.32 & 1,619.30 & 0.94 \\
        Best-of-4+RM & 83.54 & 2,408.62 & 1.15 & 73.07 & 1,294.26 & 0.72 & 42.54 & 3,017.06 & 1.44 \\
        \textbf{\textsc{TipCoder}-RL+RM} & 88.21 & 2,812.13 & 0.99 & 73.81 & 1,210.38 & 0.73 & 43.89 & 2,545.53 & 1.15 \\
        Best-of-8+RM & 84.15 & 4,610.02 & 1.94 & 73.54 & 2,468.28 & 1.20 & 44.56 & 5,828.02 & 2.48 \\
        Best-of-16+RM & 84.15 & 8,980.98 & 3.41 & 74.60 & 4,813.69 & 2.11 & 46.05 & 11,448.37 & 4.37 \\
        \bottomrule
    \end{tabular}
    }
\end{table*}

On HumanEval+, \textsc{TipCoder}-RL reaches 88.21 Pass@1 with 2,812 tokens and 0.99 seconds, compared with 84.15, 4,610 tokens, and 1.94 seconds for Best-of-8. On MBPP+, it reaches 73.81 with 1,210 tokens, while Best-of-8 reaches 73.54 with 2,468 tokens; Best-of-16 improves to 74.60 using 4,814 tokens. On BigCodeBench-full, \textsc{TipCoder}-RL reaches 43.89 with 2,546 tokens, exceeding Best-of-4 at 42.54 with 3,017 tokens. Best-of-8 reaches 44.56 with 5,828 tokens. Overall, \textsc{TipCoder}-RL exceeds Best-of-8 on HumanEval+ and MBPP+ with fewer tokens and falls between Best-of-4 and Best-of-8 on BigCodeBench-full.

\subsection{Systematic Analysis of Generated Tips}
\label{sec:tip_analysis}

We analyze the tips selected by \texttt{AceCodeRM-32B} across three seeds, four target Code LLMs, and HumanEval+/MBPP+, totaling 6,504 evaluated instances. The RM selects the guided branch 2,187 times: 310 selections repair a base failure, 117 corrupt a correct base solution, and 1,760 leave execution correctness unchanged. The selected tips therefore yield a net gain of 193 corrected instances, equivalent to 2.97 micro-averaged Pass@1 points.

\begin{table*}[t]
    \centering
    \small
    \setlength{\tabcolsep}{6pt}
    \caption{Outcome-conditioned analysis of generated-tip categories. Positive/negative denote repairing a base failure/corrupting a correct base solution.}
    \label{tab:tip_categories}
    \begin{tabular}{lrrrr}
        \toprule
        \textbf{Selected-tip category} & \textbf{Positive} & \textbf{Negative} & \textbf{Net} & \textbf{Positive rate among changed} \\
        \midrule
        Semantic/constraint clarification & 134 & 25 & +109 & 84.3\% \\
        Boundary/edge-case handling & 56 & 17 & +39 & 76.7\% \\
        Index/order/output alignment & 56 & 21 & +35 & 72.7\% \\
        Algorithmic strategy/invariant & 32 & 15 & +17 & 68.1\% \\
        Type/parsing/API guidance & 18 & 18 & 0 & 50.0\% \\
        Numerical/formula/precision & 14 & 21 & -7 & 40.0\% \\
        \midrule
        \textbf{Total} & \textbf{310} & \textbf{117} & \textbf{+193} & \textbf{72.6\%} \\
        \bottomrule
    \end{tabular}
\end{table*}
Semantic/constraint clarification is the strongest source of repairs, accounting for 43.2\% of all positive changes and a net gain of 109 instances. Boundary cases, index/order/output alignment, and algorithmic guidance are also net-positive. Type/parsing/API guidance is balanced, whereas numerical/formula/precision guidance is net-negative. For example, the tip for HumanEval/26 clarifies duplicate-removal semantics and repairs a failure, while the tip for HumanEval/7 introduces an unsupported case-insensitive-matching assumption and breaks a correct solution. These examples also show that the selector is useful but imperfect.

\section{Related Work}

\paragraph{Iterative Refinement with Code LLMs} Iterative refinement constitutes a post-hoc scaling strategy where models systematically correct outputs based on feedback. Originating in general reasoning domains approaches like Self-Refine~\cite{madaan2023self} and Reflexion~\cite{shinn2023reflexion} prompted models to critique their own generations using intrinsic knowledge. In the context of code generation this methodology has evolved to leverage objective execution signals such as compiler errors to enhance debugging capabilities~\cite{zheng2024opencodeinterpreter}. Contemporary research further internalizes this refinement through training objectives exemplified by SCoRe~\cite{kumartraining} which optimizes the entire generation trajectory via reinforcement learning and ReflectionCoder~\cite{ren2025reflectioncoder} which distills knowledge from execution-based reflection sequences. CTRL~\cite{xie2025teaching} trains LLMs to critique without human supervision, enabling them to supervise stronger models and achieve test-time scaling through iterative critique-revisions. However, these methods necessitate deployed execution environments and explicit feedback loops introducing significant workflow complexity. \textsc{TipCoder} addresses this by distilling the foresight from iterative repair into a lightweight proactive tip achieving refinement benefits in a single inference pass without runtime execution.

\paragraph{Prompt Optimization and Scaling} Prompt optimization functions as inference-time scaling by refining input queries to elicit superior responses. General domain research treats instruction generation as a search problem where models iteratively propose and score candidate prompts~\cite{zhou2022large}. Recent frameworks like DIPPER~\cite{hu2025dipper} further demonstrate that enforcing prompt diversity effectively unlocks varied reasoning paths. These advancements establish that allocating compute to explore the instruction space yields predictable gains comparable to increasing model size~\cite{snell2025scaling}. Despite this success, systematic instruction-level scaling remains underexplored in code generation where methods typically rely on stochastic solution sampling. \textsc{TipCoder} addresses this gap by introducing \textit{Test-time Instruction Proposing} to code generation. We learn a proposer policy to efficiently search for reasoning tips and convert inference compute into correctness gains through instruction-space optimization.

\section{Conclusion}
We presented \textsc{TipCoder}, a black-box test-time framework that explores the instruction space as a complement to solution-space sampling. By training a Proposer via SFT and RL to generate auxiliary instructions, \textsc{TipCoder} exposes useful alternative candidates without requiring target-model parameters, gradients, or runtime execution feedback. Across the evaluated Python code-generation benchmarks, the same trained Proposer transfers to four target backbones and is competitive with larger sampling budgets under RM-based selection. Future work includes training with multiple rollout environments, improving post-hoc selection, and evaluating multilingual and repository-level code generation.

\section*{Limitations}
\textsc{TipCoder} has three main limitations. First, it introduces additional inference cost by generating an auxiliary tip and a tip-guided code candidate before reward-model selection. Second, its realized gains depend on the quality of the post-hoc Reward Model: learned tips may expose better candidates, but an imperfect selector may fail to choose them. Third, although the Proposer transfers across multiple target Code LLMs in our experiments, it is trained with a specific rollout environment and debugging-data construction pipeline, so its guidance may still reflect the error patterns of the training environment. Extending \textsc{TipCoder} to broader model families, programming languages, and repository-level tasks remains future work.


\bibliography{custom}

\appendix

\section{Best-of-N Scaling Across Target Backbones}
\label{app:best_of_n_mbpp}

\begin{figure}[h]
    \centering
    \includegraphics[width=\linewidth]{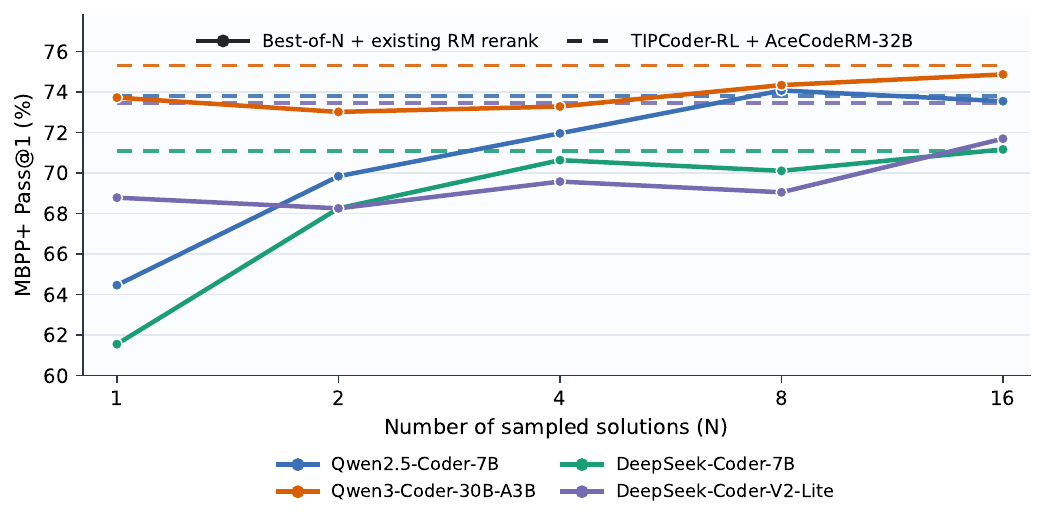}
    \caption{
    Comparison between solution-space scaling and instruction-guided scaling on MBPP+.
    Solid curves show Best-of-$N$ sampling with RM-based reranking for $N\in\{1,2,4,8,16\}$.
    Dashed horizontal lines show \textsc{TipCoder}-RL with \texttt{AceCodeRM-32B}.
    }
    \label{fig:best_of_n_mbpp}
\end{figure}

Figure~\ref{fig:best_of_n_mbpp} extends Table~\ref{tab:cost_normalized_bon} by comparing \textsc{TipCoder}-RL with Best-of-$N$ sampling on MBPP+ across all four target Code LLM backbones. \textsc{TipCoder}-RL remains competitive with the Best-of-4-to-8 range across the evaluated backbones.

\section{Discussion about Post-hoc Verification Strategy}
The post-hoc verification strategy is favored over pre-hoc refusal to maintain proposer diversity and adaptability. Training a refusal mechanism often leads to reward hacking, resulting in conservative policies that suppress high-value guidance. Furthermore, since problem difficulty is model-dependent, the proposer cannot accurately predict failure without access to the generator's internal state. By utilizing a specialized RM to filter candidates, \textsc{TipCoder} implements a dynamic, model-adaptive scaling strategy that maximizes the available potential range without parameter intervention.

\section{Hyperparameter Settings}

\label{app:hyperparams}

To ensure the reproducibility of our results, we detail the hyperparameter configurations used in our experiments. Our training pipeline consists of two stages: Supervised Fine-Tuning (SFT) and Reinforcement Learning (RL).

In the SFT stage, we fine-tuned the \texttt{Qwen3-4B} model using the Adam optimizer with state offloading to optimize memory efficiency. We employed a maximum sequence length of 4096 to accommodate long code contexts and set the learning rate to $3\text{e-}5$.

For the RL stage, we utilized the OpenRLHF framework to implement the GRPO algorithm. To maintain stability during the alignment process, we used a lower learning rate of $1\text{e-}6$ for the actor model and initialized the KL divergence coefficient at 1e-3. We employed group normalization for advantage estimation and applied reward normalization to stabilize the training scale. Additionally, a length penalty of 1e-5 was applied to the RL reward function to discourage verbosity. We conducted all experiments on a computational node equipped with 4 NVIDIA A800 GPUs. To efficiently compute the KL divergence constraint during the GRPO phase we allocated one dedicated GPU for the reference model inference while the remaining three GPUs were utilized for optimizing the policy model. Consequently the global batch sizes were specifically configured to be divisible by 3 to ensure balanced data parallelism across the active training devices.

Table~\ref{tab:hyperparams} presents the comprehensive list of hyperparameters.

\begin{table}[h]
    \centering
    \small
    \renewcommand{\arraystretch}{1.1} 
    \setlength{\tabcolsep}{4pt}       

    \begin{tabular}{lr}
        \toprule
        \textbf{Hyperparameter} & \textbf{Value} \\
        \midrule
        \multicolumn{2}{c}{\textit{\textbf{Common Settings}}} \\
        \midrule
        Base Model & Qwen3-4B-Instruct-2507 \\
        Precision & BF16 \\
        Zero Stage & 2 \\
        Gradient Checkpointing & True \\

        \midrule
        \multicolumn{2}{c}{\textit{\textbf{Stage 1: Supervised Fine-Tuning}}} \\
        \midrule
        Optimizer & Adam (Offload) \\
        Learning Rate & $3\times 10^{-5}$ \\
        Max Sequence Length & 4096 \\
        Micro Batch Size & 4 \\
        Max Epochs & 1 \\
        Logging Steps & 1 \\

        \midrule
        \multicolumn{2}{c}{\textit{\textbf{Stage 2: Reinforcement Learning (GRPO)}}} \\
        \midrule
        Framework & OpenRLHF \\
        Actor Learning Rate & $1\times 10^{-6}$ \\
        Initial KL Coefficient ($\beta$) & $1\times 10^{-3}$ \\
        Advantage Estimator & GRPO \\
        Reward Normalization & True \\
        Reward Length Penalty & $1\times 10^{-5}$ \\
        \midrule
        \textit{Batch Configuration} & \\
        \hspace{1em}Train Batch Size & 96 \\
        \hspace{1em}Rollout Batch Size & 192 \\
        \hspace{1em}Micro Train Batch Size & 8 \\
        \hspace{1em}Micro Rollout Batch Size & 8 \\
        \hspace{1em}Samples per Prompt ($N$) & 8 \\
        \midrule
        \textit{Generation Specs} & \\
        \hspace{1em}Prompt Max Length & 4096 \\
        \hspace{1em}Generate Max Length & 4096 \\
        \midrule
        \textit{Training Loop} & \\
        \hspace{1em}Max Epochs & 1 \\
        \hspace{1em}Num Episodes & 1 \\
        \bottomrule
    \end{tabular}
    \caption{Comprehensive hyperparameter settings for SFT and RL stages.}
    \label{tab:hyperparams}
\end{table}

\section{Case Study}

Figure~\ref{fig:case_study} presents a representative example where the original problem instruction is ambiguous about whether the input lists have equal lengths.
The base model assumes equal lengths and iterates over the first list, which can cause an index error when the other lists are shorter.
\textsc{TipCoder} generates tips that explicitly warn about variable-length inputs and index alignment.
Conditioned on these tips, the generator uses the minimum list length and handles the edge case correctly.
This example illustrates the main role of the Proposer: it does not provide the solution code, but surfaces problem-specific constraints and pitfalls that steer the generator away from a likely failure mode.
\begin{figure*}[htbp]
    \centering
    \lstset{
        language=Python,
        basicstyle=\ttfamily\scriptsize, 
        keywordstyle=\color{blue!70!black}\bfseries,
        commentstyle=\color{green!50!black}\itshape,
        stringstyle=\color{orange!90!black},
        breaklines=true,
        frame=tb, 
        rulecolor=\color{black!20},
        backgroundcolor=\color{gray!5},
        aboveskip=0.5em,
        belowskip=0.5em,
        numbers=none
    }

    \noindent\rule{\linewidth}{1.5pt} 
    \vspace{0.5em}

    \begin{minipage}{0.98\linewidth}
        \textbf{\large \textsc{Case Study: Handling Implicit Constraints (MBPP/142)}}
        \vspace{0.5em}

        \textbf{[User Instruction]} \\
        \vspace{0.3em}
        \textit{Write a function to count number items that are identical in the same position of three given lists.}

        \vspace{0.5em}
        \small{\textcolor{gray}{Note: The prompt is ambiguous regarding whether the three lists have equal lengths.}}
    \end{minipage}

    \vspace{0.8em}
    \hrule height 0.5pt 
    \vspace{0.8em}

    \begin{minipage}{0.98\linewidth}
        \textbf{\textsc{Generated Tips (TipCoder):}}
        \vspace{0.3em}

        \textbf{1. Index Alignment:} Ensure that you compare elements at the same index across all lists. If the lists are of unequal length, you must handle cases where one list ends before another.

        \textbf{2. Variable Lengths:} Always check if the lists are empty or have different lengths before attempting to iterate. Avoid accessing indices that do not exist.

        \textbf{3. Edge Cases:} Handle cases where the lists are empty or have different lengths gracefully.
    \end{minipage}

    \vspace{0.8em}
    \hrule height 0.5pt 
    \vspace{0.8em}

    \begin{minipage}{0.48\linewidth}
        \textbf{\textsc{\textcolor{red!70!black}{Baseline (Fail)}}}
        \begin{lstlisting}[title={\itshape Logic: Iterates by List 1}]
def count_samepair(list1, list2, list3):
    count = 0
    # DANGER: Implicitly assumes equal lengths
    # Will crash if list2/list3 are shorter
    for i in range(len(list1)):
        if list1[i] == list2[i] == list3[i]:
            count += 1
    return count
        \end{lstlisting}
        \footnotesize{\textcolor{red!70!black}{\textbf{Analysis:} The model assumes fixed lengths, leading to potential \texttt{IndexError}.}}
    \end{minipage}
    \hfill 
    \vline width 0.5pt 
    \hfill
    \begin{minipage}{0.48\linewidth}
        \textbf{\textsc{\textsc{\textcolor{green!40!black}{\textsc{TipCoder} (Success)}}}}
        \begin{lstlisting}[title={\itshape Logic: Robust Length Handling}]
def count_samepair(list1, list2, list3):
    # Handle edge cases
    if not list1 or not list2 or not list3:
        return 0

    # SAFE: Use minimum length
    min_len = min(len(list1), len(list2), len(list3))

    count = 0
    for i in range(min_len):
        if list1[i] == list2[i] == list3[i]:
            count += 1
    return count
        \end{lstlisting}
        \footnotesize{\textcolor{green!40!black}{\textbf{Analysis:} Guided by the tip, the model uses \texttt{min()} to prevent crashes.}}
    \end{minipage}

    \vspace{1em}
    \noindent\rule{\linewidth}{1.5pt} 

    \caption{Comparison of code generation with and without \textsc{TipCoder}. The baseline model fails to account for variable list lengths due to prompt ambiguity, while \textsc{TipCoder} proactively generates warnings about "Variable Lengths", leading to a robust solution.}
    \label{fig:case_study}
\end{figure*}

\section{Prompt Templates}
\label{sec:appendix_prompts}
\begin{figure}[!ht]
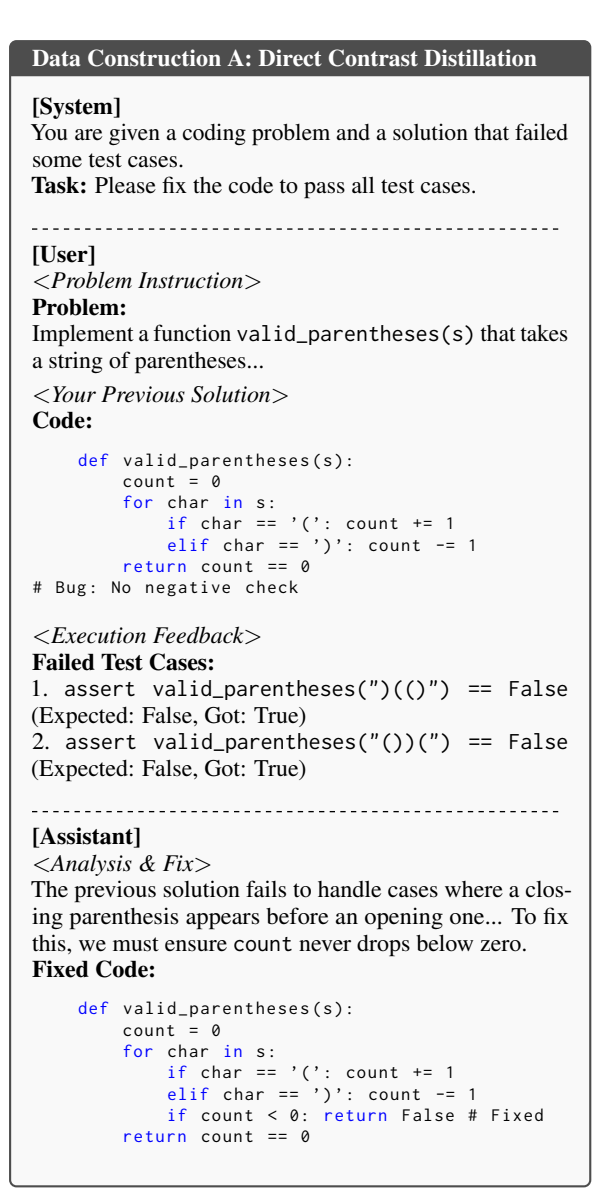

    \centering
    \begin{tcolorbox}[
        colback=white!95!gray,
        colframe=gray!60!black,
        title=\textbf{Data Construction A: Direct Contrast Distillation},
        fonttitle=\bfseries\small,
        boxrule=0.8pt,
        arc=2pt,
        left=5pt, right=5pt, top=5pt, bottom=5pt
    ]
    \small
    \textbf{[System]} \\
    You are given a coding problem and a solution that failed some test cases. \\
    \textbf{Task:} Please fix the code to pass all test cases.

    \vspace{0.2cm}
    \noindent\hdashrule[0.5ex]{\linewidth}{0.5pt}{2pt}

    \textbf{[User]} \\
    \textit{$<$Problem Instruction$>$} \\
    \textbf{Problem:} \\
    Implement a function \texttt{valid\_parentheses(s)} that takes a string of parentheses...

    \vspace{0.3em}
    \textit{$<$Your Previous Solution$>$} \\
    \textbf{Code:}
    \begin{lstlisting}[language=Python, basicstyle=\ttfamily\scriptsize, keywordstyle=\color{blue}, frame=none, numbers=none]
    def valid_parentheses(s):
        count = 0
        for char in s:
            if char == '(': count += 1
            elif char == ')': count -= 1
        return count == 0  # Bug: No negative check
    \end{lstlisting}

    \vspace{0.3em}
    \textit{$<$Execution Feedback$>$} \\
    \textbf{Failed Test Cases:} \\
    1. \texttt{assert valid\_parentheses(")(()") == False} \quad (Expected: False, Got: True) \\
    2. \texttt{assert valid\_parentheses("())(") == False} \quad (Expected: False, Got: True)

    \vspace{0.2cm}
    \noindent\hdashrule[0.5ex]{\linewidth}{0.5pt}{2pt}

    \textbf{[Assistant]} \\
    \textit{$<$Analysis \& Fix$>$} \\
    The previous solution fails to handle cases where a closing parenthesis appears before an opening one...
    To fix this, we must ensure \texttt{count} never drops below zero.

    \textbf{Fixed Code:}
    \begin{lstlisting}[language=Python, basicstyle=\ttfamily\scriptsize, keywordstyle=\color{blue}, frame=none, numbers=none]
    def valid_parentheses(s):
        count = 0
        for char in s:
            if char == '(': count += 1
            elif char == ')': count -= 1
            if count < 0: return False # Fixed
        return count == 0
    \end{lstlisting}
    \end{tcolorbox}
    \caption{The prompt template used to distill tips by directly comparing an incorrect solution with a correct one. This simulates the teacher identifying the gap between the student's attempt and the ground truth.}
    \label{fig:w4_prompt_template}
    \end{figure}

\begin{figure}[ht]
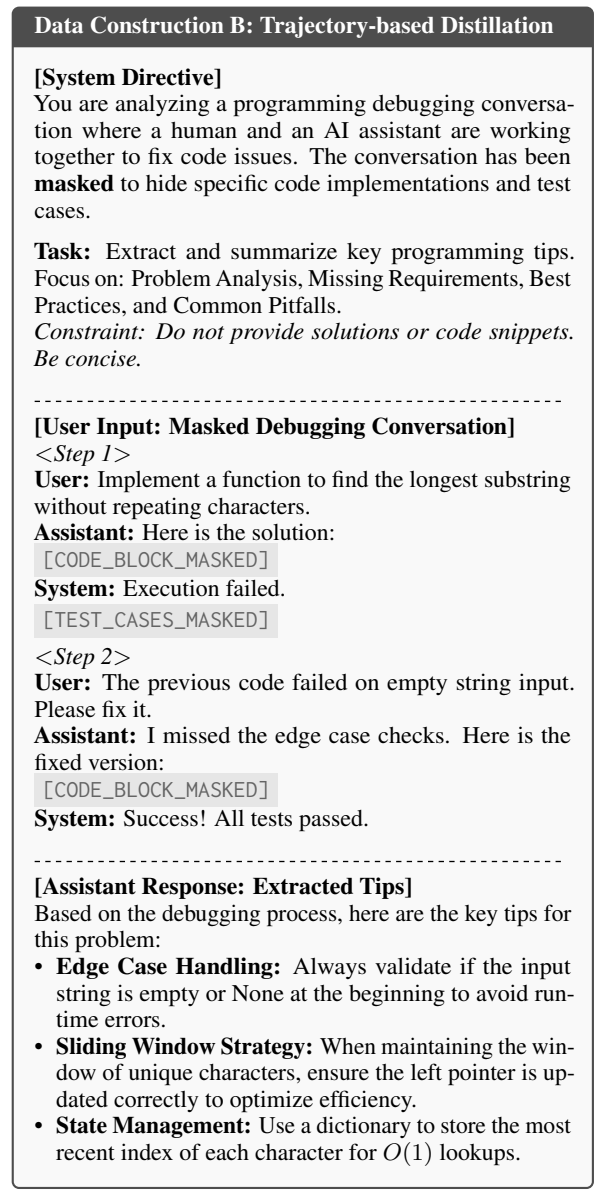

\centering
\definecolor{boxframe}{RGB}{180, 180, 180} 
\definecolor{boxbg}{RGB}{250, 250, 250}    
\definecolor{maskbg}{RGB}{230, 230, 230}   
\definecolor{masktext}{RGB}{100, 100, 100} 

\begin{tcolorbox}[
    colback=white!95!gray,
    colframe=gray!60!black,
    title=\textbf{Data Construction B: Trajectory-based Distillation},
    fonttitle=\bfseries\small,
    boxrule=0.8pt,
    arc=2pt,
    left=5pt, right=5pt, top=5pt, bottom=5pt
]
\small

\textbf{[System Directive]} \\
You are analyzing a programming debugging conversation where a human and an AI assistant are working together to fix code issues. The conversation has been \textbf{masked} to hide specific code implementations and test cases.

\vspace{0.2cm}
\textbf{Task:} Extract and summarize key programming tips. Focus on: Problem Analysis, Missing Requirements, Best Practices, and Common Pitfalls. \\
\textit{Constraint: Do not provide solutions or code snippets. Be concise.}

\vspace{0.2cm}
\noindent\hdashrule[0.5ex]{\linewidth}{0.5pt}{2pt}

\textbf{[User Input: Masked Debugging Conversation]} \\
\textit{$<$Step 1$>$} \\
\textbf{User:} Implement a function to find the longest substring without repeating characters. \\
\textbf{Assistant:} Here is the solution: \\
\colorbox{maskbg}{\texttt{\textcolor{masktext}{[CODE\_BLOCK\_MASKED]}}} \\
\textbf{System:} Execution failed. \\
\colorbox{maskbg}{\texttt{\textcolor{masktext}{[TEST\_CASES\_MASKED]}}}

\vspace{0.4em}
\textit{$<$Step 2$>$} \\
\textbf{User:} The previous code failed on empty string input. Please fix it. \\
\textbf{Assistant:} I missed the edge case checks. Here is the fixed version: \\
\colorbox{maskbg}{\texttt{\textcolor{masktext}{[CODE\_BLOCK\_MASKED]}}} \\
\textbf{System:} Success! All tests passed.

\vspace{0.2cm}
\noindent\hdashrule[0.5ex]{\linewidth}{0.5pt}{2pt}
\textbf{[Assistant Response: Extracted Tips]} \\
Based on the debugging process, here are the key tips for this problem:

\begin{itemize}[leftmargin=*, nosep]
    \item \textbf{Edge Case Handling:} Always validate if the input string is empty or None at the beginning to avoid runtime errors.
    \item \textbf{Sliding Window Strategy:} When maintaining the window of unique characters, ensure the left pointer is updated correctly to optimize efficiency.
    \item \textbf{State Management:} Use a dictionary to store the most recent index of each character for $O(1)$ lookups.
\end{itemize}

\end{tcolorbox}
\caption{Trajectory-based Distillation. The template employs a masking strategy to abstract away specific implementation details from multi-turn debugging logs, forcing the model to synthesize high-level optimization tips based on the execution feedback loop.}
\label{fig:w4_tips_prompt}
\end{figure}

\begin{figure}[ht]
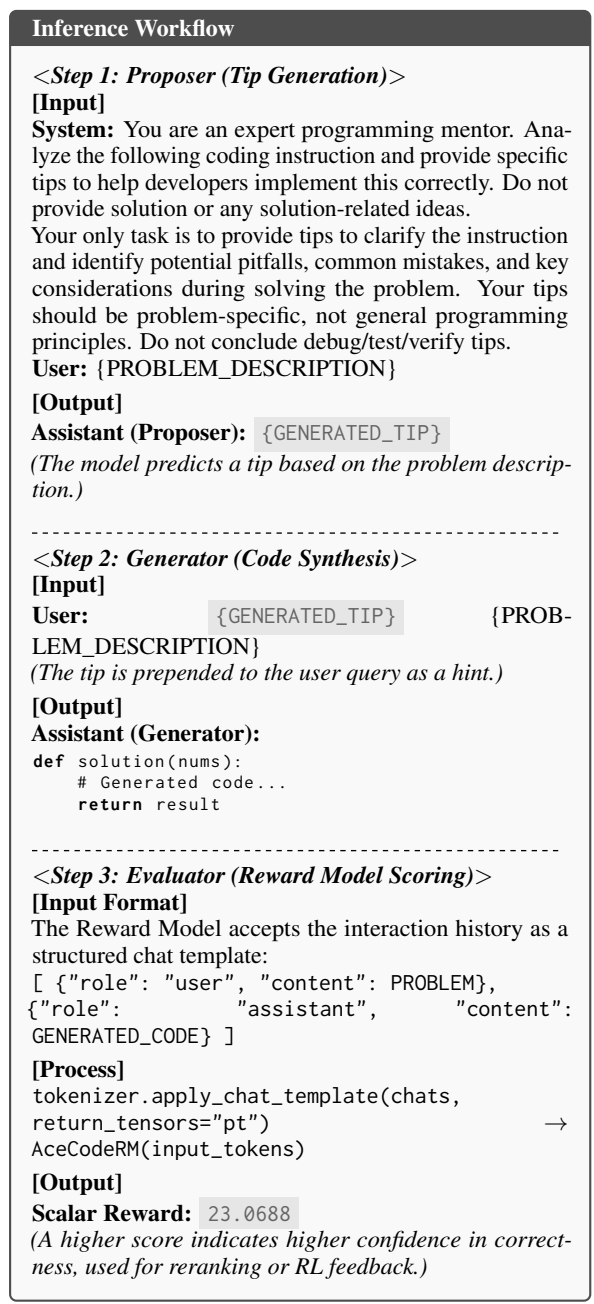

\centering
\definecolor{boxframe}{RGB}{180, 180, 180}
\definecolor{boxbg}{RGB}{250, 250, 250}
\definecolor{maskbg}{RGB}{230, 230, 230}
\definecolor{masktext}{RGB}{100, 100, 100}

\begin{tcolorbox}[
    colback=white!95!gray,
    colframe=gray!60!black,
    title=\textbf{Inference Workflow},
    fonttitle=\bfseries\small,
    boxrule=0.8pt,
    arc=2pt,
    left=5pt, right=5pt, top=5pt, bottom=5pt
]
\small

\textbf{\textit{$<$Step 1: Proposer (Tip Generation)$>$}} \\
\textbf{[Input]} \\
\textbf{System:}
You are an expert programming mentor. Analyze the following coding instruction and provide specific tips to help developers implement this correctly. Do not provide solution or any solution-related ideas. \\
Your only task is to provide tips to clarify the instruction and identify potential pitfalls, common mistakes, and key considerations during solving the problem. Your tips should be problem-specific, not general programming principles. Do not conclude debug/test/verify tips. \\
\textbf{User:} \{PROBLEM\_DESCRIPTION\}

\vspace{0.3em}
\textbf{[Output]} \\
\textbf{Assistant (Proposer):} \colorbox{maskbg}{\texttt{\textcolor{masktext}{\{GENERATED\_TIP\}}}} \\
\textit{(The model predicts a tip based on the problem description.)}

\vspace{0.2cm}
\noindent\hdashrule[0.5ex]{\linewidth}{0.5pt}{2pt}

\textbf{\textit{$<$Step 2: Generator (Code Synthesis)$>$}} \\
\textbf{[Input]} \\
\textbf{User:} \colorbox{maskbg}{\texttt{\textcolor{masktext}{\{GENERATED\_TIP\}}}} \quad \{PROBLEM\_DESCRIPTION\} \\
\textit{(The tip is prepended to the user query as a hint.)}

\vspace{0.3em}
\textbf{[Output]} \\
\textbf{Assistant (Generator):}
\begin{lstlisting}[language=Python, basicstyle=\ttfamily\scriptsize, frame=none, numbers=none, aboveskip=2pt, belowskip=2pt]
def solution(nums):
    # Generated code...
    return result
\end{lstlisting}

\vspace{0.2cm}
\noindent\hdashrule[0.5ex]{\linewidth}{0.5pt}{2pt}

\textbf{\textit{$<$Step 3: Evaluator (Reward Model Scoring)$>$}} \\
\textbf{[Input Format]} \\
The Reward Model accepts the interaction history as a structured chat template: \\
\texttt{[ \{"role": "user", "content": PROBLEM\}, \\ \{"role": "assistant", "content": GENERATED\_CODE\} ]}

\vspace{0.3em}
\textbf{[Process]} \\
\texttt{tokenizer.apply\_chat\_template(chats, return\_tensors="pt")} $\rightarrow$ \texttt{AceCodeRM(input\_tokens)}

\vspace{0.3em}
\textbf{[Output]} \\
\textbf{Scalar Reward:} \colorbox{maskbg}{\texttt{\textcolor{masktext}{23.0688}}} \\
\textit{(A higher score indicates higher confidence in correctness, used for reranking or RL feedback.)}

\end{tcolorbox}
\caption{The complete inference pipeline. In Step 1, the Proposer predicts a tip. In Step 2, the Generator synthesizes code using the tip. Finally, in Step 3, the AceCodeRM evaluates the quality of the (Problem, Code) pair to assigning a scalar reward.}
\label{fig:inference_pipeline}
\end{figure}

We present the comprehensive prompt specifications utilized throughout the \textsc{TipCoder} framework. The templates are categorized into offline data construction protocols and online inference workflows. For data construction we design two distinct mechanisms where the first employs direct contrast between incorrect and correct solutions to distill static tips while the second leverages multi-turn debugging trajectories to extract execution-based reasoning logic. The inference process follows a structured three-stage pipeline involving a Proposer for tip generation and a Generator for code synthesis followed by an Evaluator for quality scoring. Visualizations of these templates are provided in Figures~\ref{fig:w4_prompt_template}, \ref{fig:w4_tips_prompt}, \ref{fig:inference_pipeline} to illustrate the input-output format and system instructions.

\section{Use of AI Assistants}
We used AI assistants to improve writing clarity and automate parts of experimental data collection. We take full responsibility for the content of this paper.

\end{document}